\documentclass{aa}
\usepackage{graphicx}

\begin{document}

\title{Absolute timing with IBIS, SPI and JEM-X aboard INTEGRAL\thanks{Based
on observations with INTEGRAL, an ESA project with instruments and science data centre
funded by ESA member states (especially the PI countries: Denmark, France, Germany, Italy, Switzerland, 
Spain), Czech Republic and Poland, and with the participation of Russia and the USA.}
}

\subtitle{Crab main-pulse arrival times in radio, X-rays and high-energy $\gamma$-rays}

\author{L. Kuiper \inst{1}
 \and W. Hermsen \inst{1}
 \and R. Walter \inst{2,3}
 \and L. Foschini \inst{4}}

\institute{SRON National Institute for Space Research, Sorbonnelaan 2, 3584 CA Utrecht, The Netherlands
 \and INTEGRAL Science Data Centre, Chemin d'Ecogia 16, CH-1290 Versoix, Switzerland
 \and Geneva Observatory, Chemin des Maillettes 51, CH-1290 Sauverny, Switzerland
 \and IASF/CNR, sezione di Bologna, via P. Gobetti 101, 40129 Bologna, Italy}
 
\offprints{L. Kuiper,\\ \email{L.M.Kuiper@sron.nl}}

\date{Received July XX / Accepted August XX}

\abstract{We have verified the absolute timing capabilities of the high-energy instruments 
aboard INTEGRAL, i.e. the 
imager IBIS, the spectrometer SPI and the X-ray monitor JEM-X. Calibration observations of the Crab, 
contemporaneous with the Rossi X-ray Timing Explorer (RXTE), have been used to measure the absolute phase 
of the main pulse of the Crab pulse profile using the same Jodrell Bank radio ephemeris. The three INTEGRAL 
instruments and RXTE give within the statistical and systematic uncertainties consistent results: The
X-ray main pulse is leading the radio pulse by $280\pm40\mu{\rm s}$. Also the shapes of the X-ray pulse 
profiles as measured by the different instruments are fully consistent with each other. 
In addition, we present
the first measurement of the absolute phase of the main pulse at $\gamma$-ray energies above 30 MeV using
data from the EGRET instrument aboard the Compton Gamma-Ray Observatory:  The $\gamma$-ray main 
pulse is leading the 
radio one by $241\pm29\mu{\rm s}$, consistent with the value for the X-ray main pulse. Comparing absolute 
arrival times at multiple frequencies gives important constraints
to models explaining the production of non-thermal emission in magnetospheres of rotation powered 
neutron stars.

}

\maketitle

\section{Introduction}
The high-energy instruments aboard INTEGRAL (Winkler et al. \cite{winkler}) are 
designed to allow timing studies of e.g. radio 
pulsars with periods down to the very short periods of millisecond pulsars. This requires very 
precise and stable timing information, particularly for studies at the high-energy 
end of the INTEGRAL window where the counting rates are low and long integration times required. 
An important aspect in timing studies is the measurement of absolute arrival times. Particularly
in multiwavelengths studies of pulsars, the comparison of absolute arrival times is an important 
diagnostic in the detailed modelling of the structures of the production sites in magnetospheres
of rotation-powered neutron stars. 

We used INTEGRAL calibration observations of the Crab pulsar
(PSR B0531+21) and 
contemporaneous observations with the Proportional Counter Array (PCA, Jahoda \cite{jahoda}) 
on the Rossi X-ray Timing Explorer (RXTE) to verify the
correct absolute timing calibration for the imager IBIS (Ubertini et al. \cite{ubertini}), 
the spectrometer SPI (Vedrenne et al. \cite{vedrenne}) and the X-ray 
monitor JEM-X (Lund et al. \cite{lund}). This was done by measuring the absolute phase of the 
main pulse of the Crab pulse profile
with the INTEGRAL instruments and the PCA. All data have been 
analyzed with the same Jodrell Bank radio ephemeris, reducing systematic effects due to e.g. 
dispersion measure variations causing systematic variations in the measured arrival times.

Finally, the arrival time of the main pulse of the Crab pulsar has been studied from radio, IR, 
optical up to X-ray wavelengths (e.g. Ray et al. \cite{ray}). We present here a first estimate 
of the $\gamma$-ray (30 MeV -- 30 GeV) radio
delay using data from EGRET (Thompson et al. \cite{thompson}) aboard the Compton Gamma-Ray Observatory (CGRO). 

\section{INTEGRAL timing resolution}

In order to stay within the limited telemetry budget of INTEGRAL, the event timing 
information measured by the instruments has been degraded
on board. The accuracy of the On Board Time (OBT) for IBIS/ISGRI (optimized for energies 
$\sim$ 15 -- 200 keV, Lebrun et al. \cite{lebrun})
is thus $61\mu{\rm s}$, for IBIS/PICsIT (optimized for energies 175 keV -- 10 MeV, Di Cocco et al. 
\cite{dicocco}) is $64\mu{\rm s}$
(for IBIS the least significant 8 bits are
cleared), for SPI $102\mu{\rm s}$ (the least significant 11 bits are cleared) and for JEM-X $122\mu{\rm s}$ 
(the readings of the Digital Front End Electronics clock).  There are additional smaller uncertainties
affecting the event timing accuracy, e.g. the OBT might have an uncertainty up to about $16\mu{\rm s}$, 
uncertainties in the orbit predictions can cause similar uncertainties and 
for pulsar studies the conversion 
to the arrival time at the Solar System Barycentre (SSB) will again add a similar error. 
These uncertainties
affect the timing of single events and cause a broadening of derived timing signatures. We estimate 
that the resulting timing resolutions for pulsar timing studies using the arrival times at the SSB
are $90\mu{\rm s}$ for IBIS, $130\mu{\rm s}$ for SPI and $150\mu{\rm s}$ for JEM-X ($3 \sigma$ accuracy). As a first check, 
we used
the Crab calibrations to compare the pulse profiles measured by the INTEGRAL instruments 
with the RXTE profile. A more accurate calibration should be performed in the
future by studying the profiles of millisecond pulsars.

\section{Observations and data}
INTEGRAL observed the Crab from 7 February 2003 (revolution 39) up to and including 27 
February 2003 (revolution
45). The spacecraft and instruments were in different modes during this calibration period. For our 
investigations we need photon-by-photon modes in which each event is time tagged. The IBIS/ISGRI 
detector, SPI and JEM-X were all in photon-by-photon 
mode during revolution 42,
roughly in the middle of the Crab observations. We used for these three instruments the data from 
revolution 42, collected at the REDU groundstation, which 
provided ample statistics.
The IBIS/PICsIT detector, unfortunately, cannot routinely 
be configured in photon-by-photon mode due to PICsIT's high count rate and the tight telemetry 
budget of INTEGRAL. 
In revolutions 39 and 40, PICsIT was part of the time commanded in photon-by-photon mode 
allowing a test of the timing capabilities.  We used these data collected at REDU.

The RXTE PCA (Jahoda \cite{jahoda}) routinely monitors the Crab pulsar, and had one observation on 13 February
2003, roughly in the middle of our Crab calibrations, and one on 27 February 2003, at the end of the INTEGRAL
observations. We retrieved these data from the High Energy As\-tro\-physics Science Archive Research 
Center (HEASARC) at NASA/Goddard Space Flight Center.

EGRET, sensitive to high-energy $\gamma$-rays from 30 MeV to 30 GeV (Thompson et al. \cite{thompson}), 
had the Crab 20
times in its field-of-view with high signal-to-noise  over a $\sim$ 7.5 year baseline starting in 
April 1991  (details of the EGRET observations of the Crab are given in Kuiper et al. \cite{kuiper}) . 
We used these data
to study the $\gamma$-ray radio delay of the main pulse.

\begin{table}
\caption{Radio ephemeris for PSR B0531+21 (monthly monitoring at Jodrell Bank) valid during the INTEGRAL Crab calibration.}
\begin{center}
\begin{tabular}{ll}
\hline\hline
\textbf{Parameter} & \textbf{Value} \\
\hline
Val. range (MJD)                            & 52\,671--52\,699 \\
$t_{0}$ (MJD; TDB)                          & 52\,685.0 \\
$\nu_{0}$ (Hz)                              & 29.8092705122612 \\
$\dot{\nu}_{0}$  ($10^{-10}$  Hz s$^{-1}$)  & $-$3.73661 \\
$\ddot{\nu}_{0}$ ($10^{-20}$ Hz s$^{-2}$)   & 2.01 \\
RMS of ToA fit                              & $23.5\mu$s \\
Phase of radio MP at zero epoch ($t_{0}$)   & 0.2838 \\
$\alpha_{2000}$                             & 05$^{\rm h}$34$^{\rm m}$31$^{\rm s}$.972 \\
$\delta_{2000}$                             & 22{\degr}00{\arcmin}52{\arcsec}.07 \\
\hline
\multicolumn{2}{c}{} \\
\end{tabular}
\end{center}
\label{tab:ephemeris}
\end{table}

\section{Timing analysis}

The Jodrell Bank radio telescope (UK) performs monthly monitoring of PSR B0531+21, providing us 
with a Crab radio ephemeris valid for INTEGRAL revolutions 39 -- 45, 
including the dates of the RXTE observations (see Table 1). 

\subsection{RXTE analysis and results}

\begin{figure*}
\
\vbox{
  \hbox{\hspace{0.5cm} \includegraphics[width=8cm]{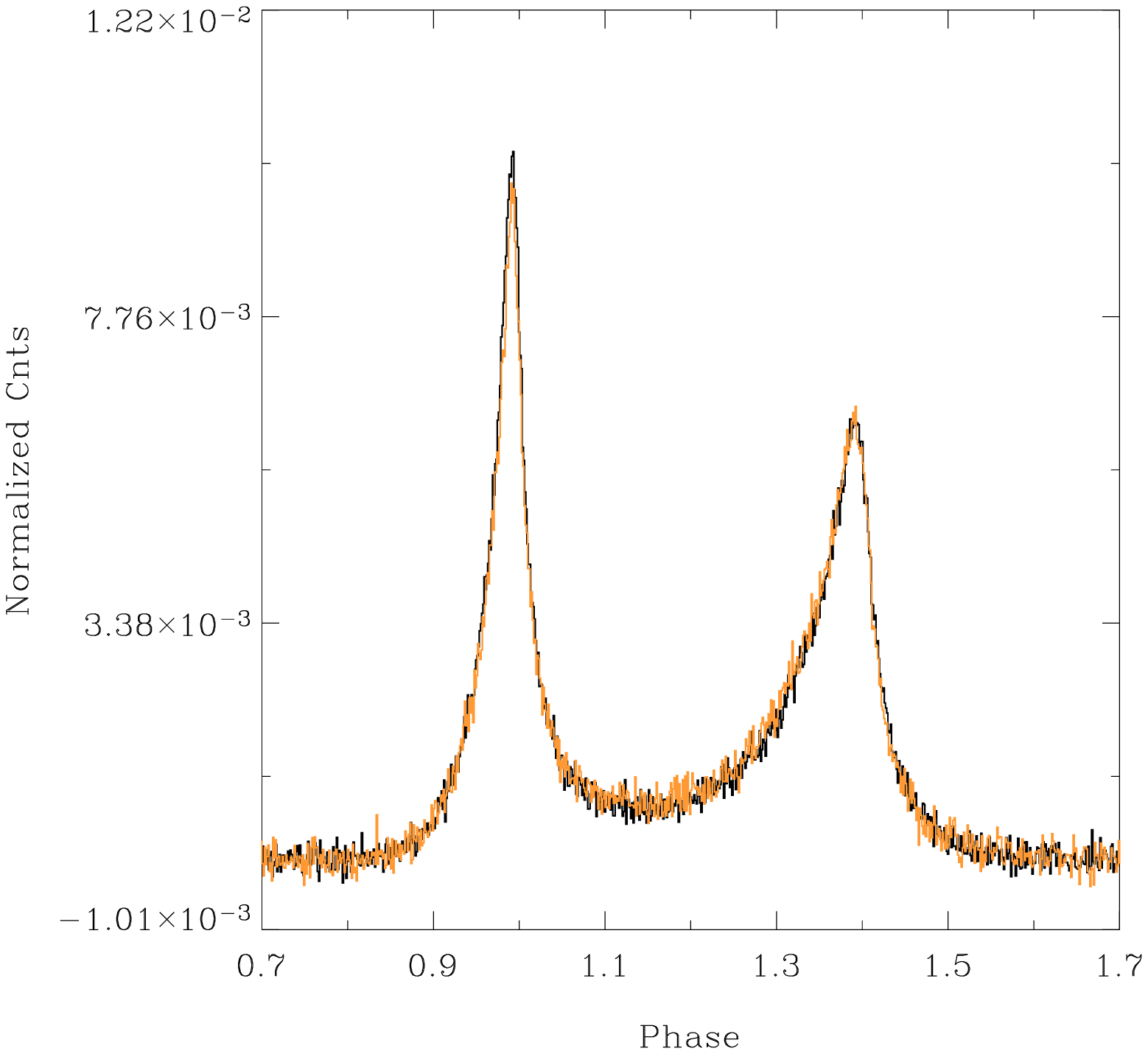} \hspace{0.75cm} \includegraphics[width=8cm]{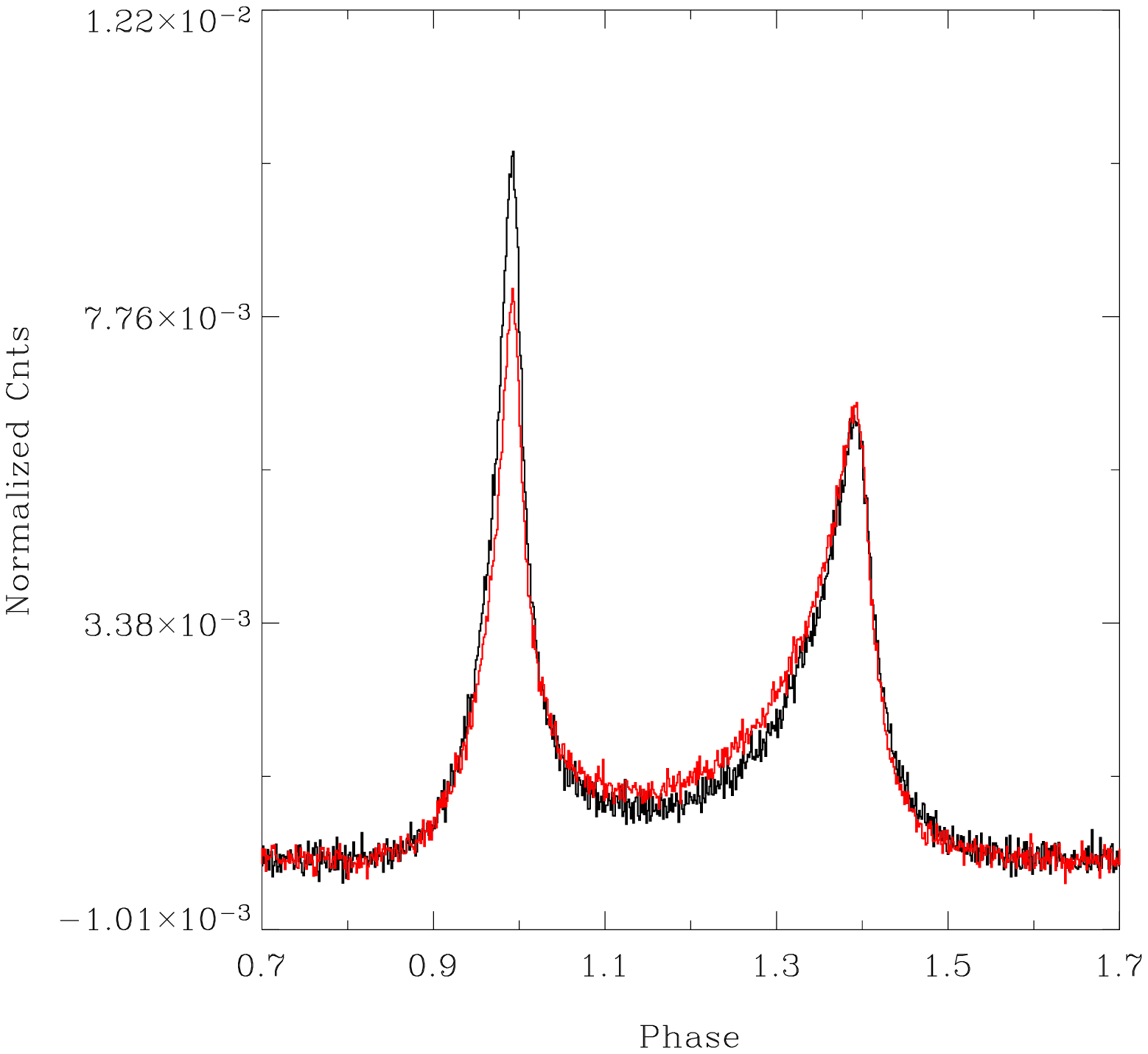} }
  \hbox{\hspace{0.5cm} \includegraphics[width=8cm]{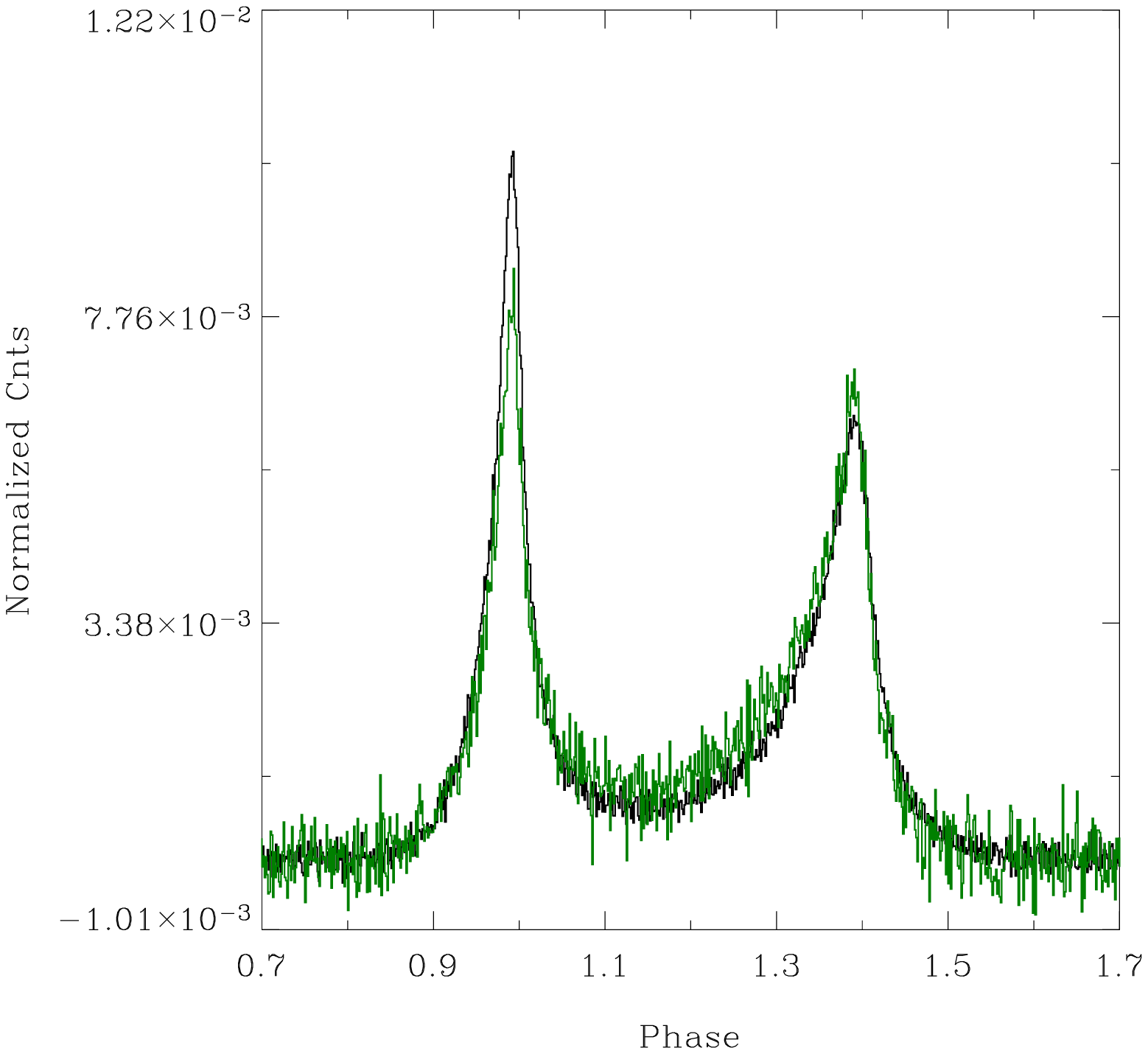} \hspace{0.75cm} \includegraphics[width=8cm]{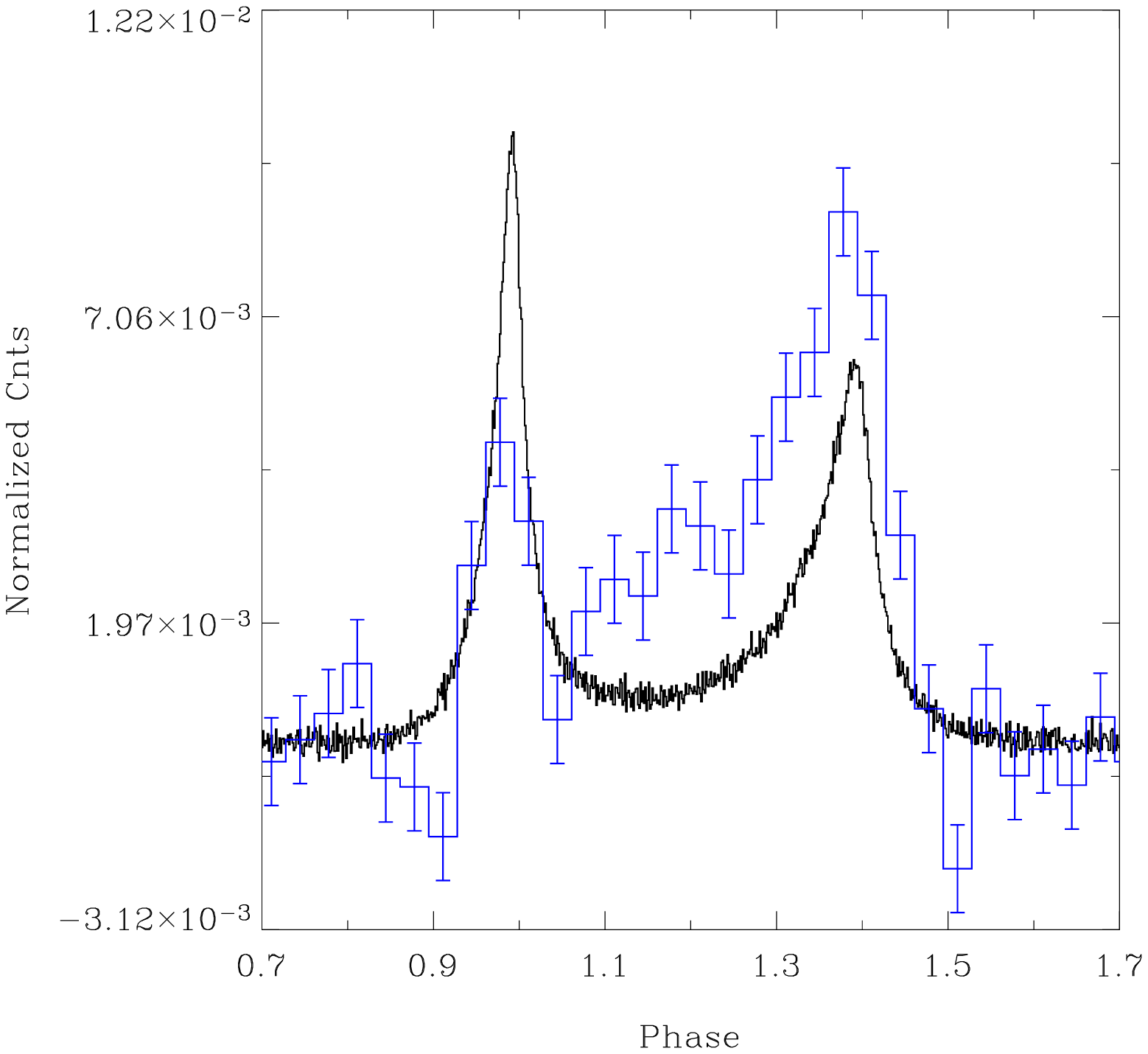}}
     }

  \caption{Comparison in absolute phase of Crab pulse profiles measured by INTEGRAL and RXTE: 
  Upper-left JEM-X (6 -- 23 keV, orange); upper-right 
  IBIS/ISGRI (20 -- 50 keV, red); lower-left SPI (20 -- 50 keV, green); lower-right 
  IBIS/PICsIT (180 -- 580 keV, blue). The RXTE 
  profile (2 -- 30 keV, black) is produced with the same ephemeris and shown in each figure.
  \label{fig-integral-rxte}}
\end{figure*}

In the timing analysis of the RXTE PCA data of 
13 and 27 February 2003 the 
barycentered event arrival times were folded with the ephemeris in Table 1, providing us with accurate 
X-ray pulse 
profiles in absolute phase (see Fig.~\ref{fig-integral-rxte}). The PCA events are time tagged with a
$1\mu{\rm s}$ accuracy  with respect to the spacecraft clock and with an absolute time accuracy of 
5 -- 8$\mu{\rm s}$ with respect to UTC.  The maximum of the main radio pulse is at phase 0.0, the fiducial
point in the radio template. For both data sets (13 and 27 February), we determined the absolute 
arrival time/phase of the main X-ray pulse by fitting 
the profile with an asymmetric Lorentzian plus a constant background. The fits were perfect 
with the X-ray maxima leading the radio
pulse by $248 \pm 7\mu{\rm s}$ and  $289 \pm 7\mu{\rm s}$, respectively. The errors are $1 \sigma$, statistical
only. The difference between the two estimates is a measure of the systematic uncertainty in the
radio parameters, e.g. due to dispersion measure variations, and is consistent with earlier RXTE estimates
of $302 \pm 67\mu{\rm s} $
(Rots et al. \cite{rots} and Tennant et al. \cite{tennant}) and also with a recent 
estimate of $\sim 300\mu{\rm s}$ using XMM-Newton data (Kirsch et al. \cite{kirsch}).

\subsection{INTEGRAL analysis and results}

For the analysis of the INTEGRAL data, we selected all time-tagged events received via 
the REDU groundstation during revolution 42. For both JEM-X monitors Pulse Height Analyzer channels 
128 -- 192 were selected, corresponding
at the time of the Crab calibration for JEM-X1 to energies 9 -- 23 keV, and JEM-X2 6 -- 15 keV (a
complementary timing study of the Crab pulsar with JEM-X is reported by Brandt et al. \cite{brandt}). 
For IBIS/ISGRI and SPI events were used with energies 20 -- 50 keV.
Finally, for IBIS/PICsIT we had $\sim$ 214 ks exposure in photon-by-photon mode 
in revolutions 39 and 40, and selected events 
with energies 180 -- 580 keV (PHA 25 -- 75). Unfortunately, the latter data set has low statistics. 

The arrival times of the events at the spacecraft are registered in On Board Time (OBT), which is converted
on ground to Terrestrial Time, expressed in INTEGRAL Julian Date (IJD)
(see Walter et al. \cite{walter}). The correlation between the OBT and IJD needs to take into
account delays on board for each of the instruments, delays within the ground stations and due to light 
travel time. A full description with correction values is given in Walter et al. (\cite{walter}). 
In this work we
applied all these corrections valid during the Crab calibrations to obtain absolute time: Offset 
added to UTC $964\mu{\rm s}$;
delays ESA REDU grounstation $-103\mu{\rm s}$ and NASA DSS-16 groundstation $42\mu{\rm s}$; delays per instrument
derived from ground calibrations,
IBIS $111\mu{\rm s}$, SPI $134\mu{\rm s}$ and JEM-X $185\mu{\rm s}$.  

The arrival times of the INTEGRAL time-tagged events were also converted to the arrival time at the
Solar System Barycentre and folded with the Crab ephemeris in Table 1. The resulting Crab profiles in
absolute phase (i.e. the main radio pulse is at phase 0.0) are 
shown in Fig.~\ref{fig-integral-rxte}, with the profiles of JEM-X1 and JEMX-2 summed in one figure. The
excellent agreement with the RXTE PCA profiles is evident for the JEM-X data. At higher energies the
INTEGRAL data display the known evolution of the Crab pulse profile with increasing energy:  1) a 
higher level of bridge emission 
between the peaks for IBIS and SPI compared to RXTE and JEM-X, 2) the relative intensity of the 
first pulse is higher
for IBIS and SPI than for RXTE and JEM-X (see e.g. Kuiper et al. \cite{kuiper}). On this scale, we do not see any broadening 
of the INTEGRAL
profiles relative to the RXTE profile. For comparison, the half-width-half-maximum of the first Crab pulse
is about $600\mu{\rm s}$.

In order the compare the absolute times more accurately, we determined for JEM-X1, JEM-X2, IBIS/ISGRI
and SPI the arrival phases of the main (first) pulses, similar to what was done for RXTE. In each case the
profile of the main pulse was fitted with an asymmetric Lorentzian plus constant background and the 
phase of the maximum was determined. The uncertainties
in the latter have a small statistical error, but there is also a systematic uncertainty which depends 
on the choice of the fit function. In order to get an estimate of the magnitude of the systematic 
uncertainties, we also made Fourier fits with 40 harmonics. Fig.~\ref{fig:crab_isgri_fits} shows 
as an example
the fits for IBIS/ISGRI. The phase difference between the two fit maxima
is $15\mu{\rm s}$, representative for the systematic uncertainties for all instruments.

For the Fourier fits, we find the following X-ray radio delays: JEM-X1 
$284 \pm 17\mu{\rm s}$, JEM-X2 $300 \pm 14\mu{\rm s}$, IBIS/ISGRI $285 \pm 12\mu{\rm s}$, SPI $265 \pm 23\mu{\rm s}$.
The errors are $1 \sigma$, statistical only. Taking the systematic uncertainties into account, 
the average value for the measured
X-ray radio delay is $\sim 280 \pm 40\mu{\rm s}$. It can be noted that, firstly, the INTEGRAL 
instruments give fully
consistent results, and, secondly, the average INTEGRAL value is fully consistent with the 
RXTE PCA values derived using the same radio ephemeris, even though 
the two X-ray measurements employ very different means of measuring time and 
satellite position
and thus have different systematic error budgets. 

Due to the tight telemetry budget, it is foreseen that IBIS/PICsIT will hardly be configured
in photon-by-photon mode. For timing studies observers can select the spectral-timing mode with 
highest time resolution of 0.97656 ms. To verify the correct performance, PICsIT was in revolutions
40 and 41 also in
spectral-timing mode for 345 ks. We produced the Crab pulse profile shown in 
Fig.~\ref{fig:picsit_spti_crab}, which has a significance of $16 \sigma$. The profile is consistent 
with those shown in 
Fig.~\ref{fig-integral-rxte}. The intense bridge emission between the peaks is expected for the selected
high energy range (band 3, 260 -- 364 keV).

{\begin{figure}[t]
  {\hspace{0.5cm} \includegraphics[width=8.0cm,height=7.0cm]{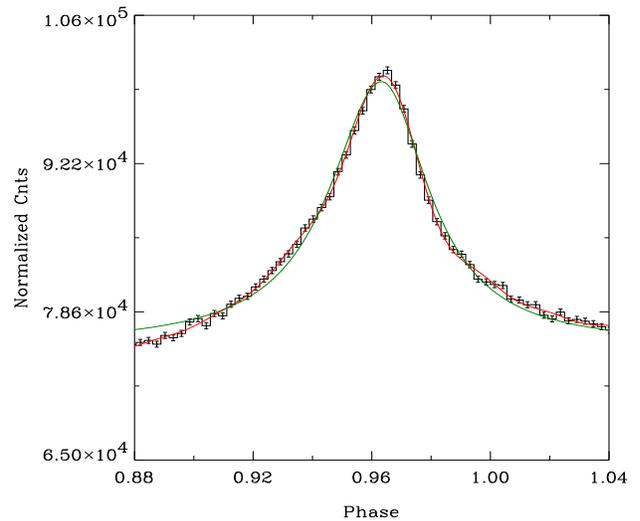}}
  \caption[]{Main pulse of the Crab profile measured by IBIS/ISGRI. Asymmetric-Lorentzian fit (green) 
  and Fourier fit with 40 harmonics (red). 
  Phase shift between the maxima of the two fits is $15\mu{\rm s}$. 
  The x-axis is not in absolute phase, the known corrections factors (see text) are not yet applied.
             \label{fig:crab_isgri_fits}}
\end{figure}}

{\begin{figure}[t]
  {\hspace{0.5cm} \includegraphics[width=8.0cm,height=7.0cm]{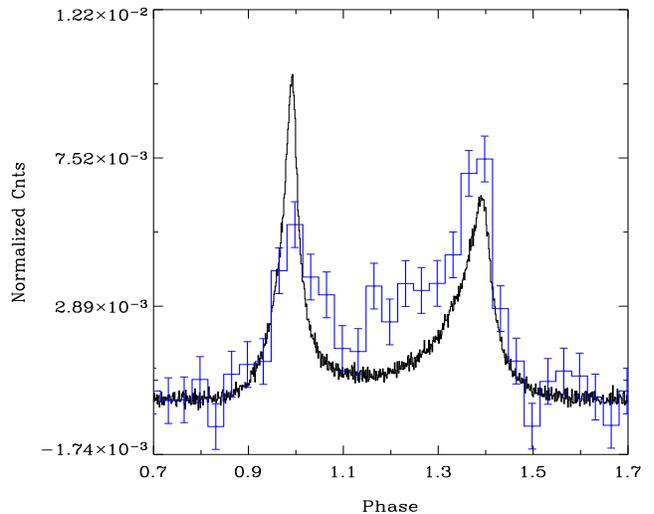}}
  \caption[]{Crab pulse profile in absolute phase measured by IBIS/PICsIT in spectral-timing mode for
             band-3 (260 -- 364 keV) using 345.2 ks exposure (blue histogramme in 30 bins). The
RXTE profile (black, 2 -- 30 keV) is shown for comparison.
             \label{fig:picsit_spti_crab}}
\end{figure}}

\subsection{Arrival time Crab main pulse at $\gamma$-ray energies above 30 MeV}

For the 20 EGRET observations with high-statistics Crab data, we constructed for all observations
the high-energy (30 MeV -- 30 GeV) pulse profiles, following Kuiper et al. \cite{kuiper}. It should be noted that 
different ephemerides had to be used over the CGRO mission lifetime, such that the different profiles
can not be directly summed in absolute phase. Due to e.g. variations in the dispersion measure,
derived arrival phases will scatter around the genuine value. The main peak
of the high-energy profile is very symmetric, because the bridge emission is very weak at these high energies. 
Therefore, we determined for each observation the arrival phase of the main pulse 
by fitting 
a symmetric Lorentzian profile plus a constant
background to the measured event distribution in a relatively broad (0.3) phase window centered 
around phase 0.
In Fig. \ref{fig:gammaarrival_crab} these arrival phases are shown
as a function of time over a $\sim 7.5$ year baseline.
The error bars on the individual datapoints are composed of a statistical contribution from the profile
fit and of a systematic component of about $100\mu{\rm s}$ ($1\sigma$) reflecting the uncertainties 
in the dispersion measure from the radio pulsar timing observations (see explanatory supplement
{\sl README.txt} in the Princeton CGRO database directory). The arrival phases have been fitted by
a constant value yielding as optimum value $-0.0072 \pm 0.0009$ for an acceptable $\chi_{\nu}^2$ value 
of 1.59 for 19 degrees of freedom.
This offset in the phase domain translates to an offset of $241\pm29\mu{\rm s}$ in the time domain and 
indicates that the high energy $\gamma$-ray photons from the main pulse also arrive earlier than the 
maximum of the main 
radio pulse. This value of $241\mu{\rm s}$ is significantly larger than the quoted CGRO absolute 
timing uncertainty of better than $100\mu{\rm s}$ and is fully consistent with the results obtained above from
the INTEGRAL and RXTE X-ray data.

{\begin{figure}[t]
  {\includegraphics[width=8.7cm,height=7.0cm]{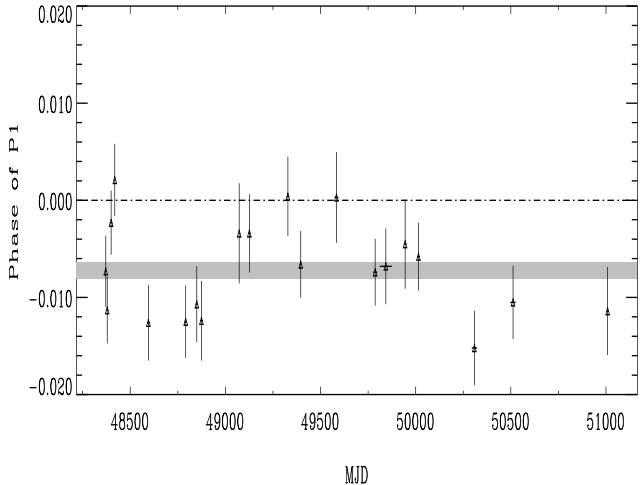}}
  \caption[]{Arrival phase of the main pulse vs time for photon energies 30 MeV -- 30 GeV derived
             from CGRO/EGRET data.
             The $\pm 1\sigma$ uncertainty interval assuming a constant arrival phase is indicated
             by the hatched region. The high-energy $\gamma$-rays arrive $241\pm29\mu{\rm s}$ earlier than
             the fiducial point in the radio template.
             \label{fig:gammaarrival_crab}}
\end{figure}}

\section{Discussion}

We have determined with all high-energy instruments aboard INTEGRAL, i.e. the 
imager IBIS, the spectrometer SPI and the X-ray monitor JEM-X, the X-ray radio delay between
the absolute arrival times of the X-ray and radio main pulses in the profile of the Crab pulsar, 
PSR B0531+21.
Within the statistical accuracies these instruments measured the same delay, on average
$\sim 280 \pm 40\mu{\rm s}$ (the error includes estimates of statistical and systematic uncertainties), the
X-ray pulse leading the radio one. 
This value is consistent with the delays we measured twice with the RXTE/PCA
during the period of the INTEGRAL Crab calibrations, namely $248 \pm 7\mu{\rm s}$ and  $289 \pm 7\mu{\rm s}$,
statistical errors only. These results show that INTEGRAL meets its specifications 
and can measure absolute arrival times with
accuracies down to about $40\mu{\rm s}$. Fig. \ref{fig:crab_isgri_radio_zoom} shows the X-ray main pulse measured by ISGRI 
for energies 20 -- 50 keV together with its radio counterpart at 812.5 MHz 
measured with the Greenbank 140 ft
telescope (Lundgren \cite{lundgren}). In addition to the phase shift, it is evident that the profile 
shapes are very different.  The narrow radio profile is shifted but contained in the phase range covered by
the X-ray profile.

{\begin{figure}[t]
  {\hspace{0.5cm} \includegraphics[width=8.0cm,height=7.0cm]{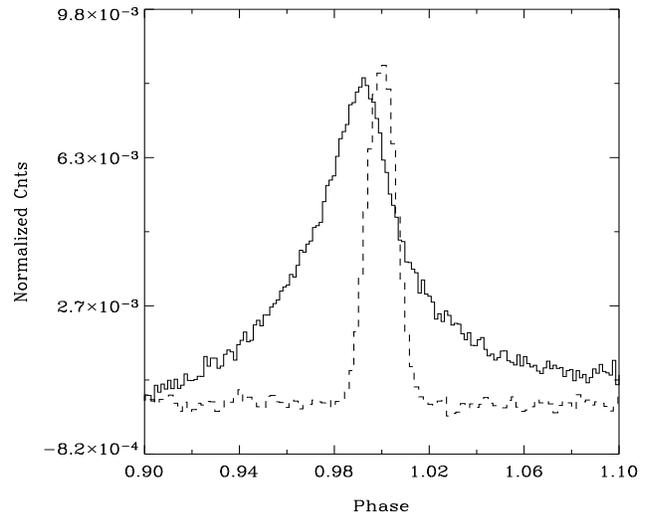}}
  \caption[]{Main pulse of the Crab profile in absolute phase as measured by 
  IBIS/ISGRI (20 -- 50 keV; full line)
  compared to the Greenbank 812.5 MHz radio profile (broken line; Lundgren \cite{lundgren}). 
             \label{fig:crab_isgri_radio_zoom}}
\end{figure}}

We also measured the delay of the arrival time of the main pulse in the high-energy (30 MeV -- 30 GeV)
$\gamma$-ray profile relative to the main radio pulse, using 20 Crab observations by CGRO/EGRET over a 
$\sim 7.5$ year baseline: $241\pm29\mu{\rm s}$. This value appears to be consistent with the delay measured
for the main X-ray pulse and is an additional constraint in the three-dimensional modelling
of the production processes and sites in the magnetospheres of rotation powered pulsars. 

In Kuiper et al.
\cite{kuiper}, presenting for the Crab a coherent high-energy picture from soft X-rays up to  high-energy 
$\gamma$-rays, it is argued that models with an outer gap are more successful in reproducing the overall
characteristics of PSR B0531+21 as measured from the radio up to the $\gamma$-ray domain, than 
polar cap models. These are the two competing model scenario's attempting to explain the
high-energy radiation from magnetized rotating neutron stars. The known large magnetic inclination angle
$\alpha$, i.e. the angle between the pulsar spin axis and magnetic moment axis, and aspect angle $\zeta$,
 i.e. the angle between the spin axis and the observer's line of sight, which are for the Crab
 both estimated from radio and optical observations to be $\sim 60^{\circ}$, cause for polar cap scenarios 
 problems to reproduce the structure of the Crab light curve (see e.g. discussion in Kuiper et al.
\cite{kuiper}).
Fig. \ref{fig:crab_radio_X_gamma} shows the phase aligned total Crab profiles
 for the energy windows we are addressing in this work (for a detailed presentation of the
 frequency-dependent behaviour of the average radio pulse profile, see Moffett $\&$ Hankins \cite{moffett}).
 Outer gap scenarios can for these large geometrical angles  $\alpha$  and $\zeta$ 
account for the measured double peaked profile with bridge emission 
in the X-ray and soft $\gamma$-ray band (e.g. Chiang \& Romani \cite{chiang}), and
are also succesful in reproducing characteristics as the observed polarization position angle swings of the 
Crab pulsar at optical
wavelengths (Romani \& Yadigaroglu \cite{romani}). The radio precursor, visible in the
430 MHz profile in Fig. \ref{fig:crab_radio_X_gamma} a, originates in the outer gap models 
from near the polar cap.      The phase shift between the
high-energy main pulse (X-rays and $\gamma$-rays) and the main radio pulse is an additional detail to
be accounted for. More recently, Cheng, Ruderman \& Zhang \cite{cheng} extended the outer magnetospheric 
gap model to a detailed three dimensional model with photon emission moving outward 
and inward from regions in outer
gaps above both poles, the gaps being limited along the azimuthal direction by $e^{\pm}$ pair production of
inward-flowing photons from the outer gaps.  In their model the main X-ray/$\gamma$-ray pulse is produced
high in the magnetosphere at 0.8 $<$ {\it $r/R_{lc}$} $<$ 1.0 with {\it $R_{lc}$} the light cylinder radius.
In order to account for the delay of the radio main pulse, the production site of radio emission in the 
 narrow outer 
gap should
be slightly shifted inwards to the neutron star with respect to the production site for the high-energy 
emisssion, but still contained in the latter. It is expected that the intensity weighting 
along the gap surface differs for different 
energy bands, but the non-thermal radio emission is likely produced in a narrower region of
the thin outer gap along the last closed field line. More detailed modelling should show whether this 
can explain 
the profiles shown in Fig. \ref{fig:crab_isgri_radio_zoom}.

{\begin{figure}[t]
  {\vspace{-1.2cm} \includegraphics[width=8.7cm,height=15.0cm]{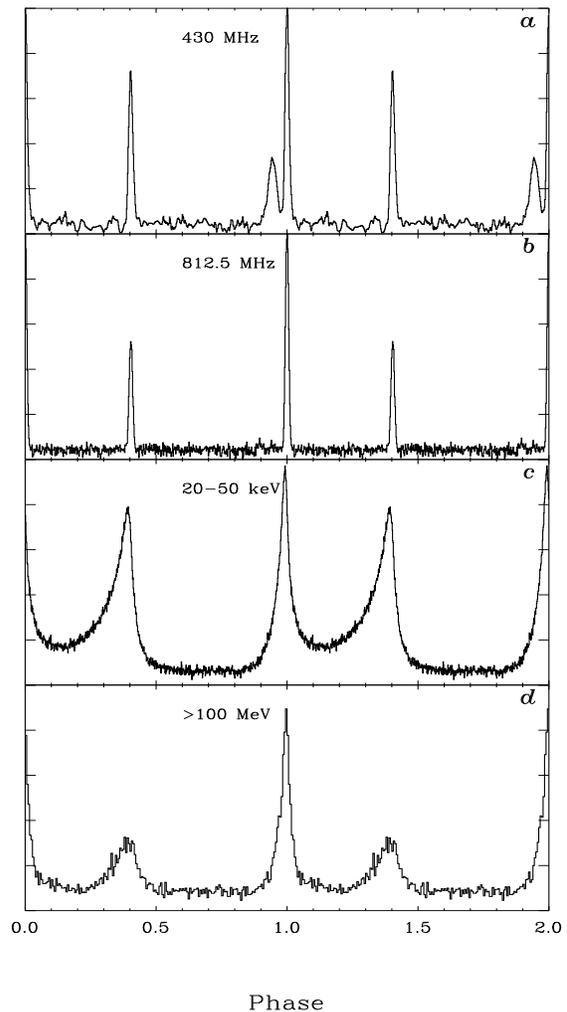}}
  \caption[]{Crab profiles in absolute phase: a) radio 430 MHz (Arecibo: 
  Moffett $\&$ Hankins \cite{moffett}), b) radio 812.5 MHz (Greenbank: Lundgren \cite{lundgren}), 
  c) X-rays 20 -- 50 keV (IBIS/ISGRI),
  d) $\gamma$-rays with energies above 100 MeV (EGRET)
  ). 
             \label{fig:crab_radio_X_gamma}}
\end{figure}}

\begin{acknowledgements}
We wish to thank the High Energy As\-tro\-physics Science Archive Research Center (HEASARC) at NASA/Goddard
Space Flight Center for maintaining its online archive service which provided the RXTE data 
used in this research. LF acknowledges financial support by the Italian Space Agency (ASI) and the 
hospitality of the ISDC during part of this work.
\end{acknowledgements}




\end{document}